\documentclass[preprint,amsmath,amssymb]{revtex4}
\usepackage{graphicx}
\usepackage{dcolumn}
\usepackage{bm}
\begin{document}
\title{Gaussian Curvature and Global effects : gravitational Aharonov-Bohm effect revisited }
\author{ M.~Nouri-Zonoz  \footnote{Electronic address:~nouri@khayam.ut.ac.ir, corresponding author}
and A.~Parvizi \footnote{Electronic
address:~a.parvizi@ut.ac.ir}} 
\address {Department of Physics, University of Tehran, North Karegar Ave., Tehran 14395-547, Iran.}
\begin{abstract}
Using the Gauss-Bonnet formula, integral of the Gaussian curvature over a 2-surface enclosed by a curve in the asymptotically flat region of a static spacetime was found to be a measure of a gravitational analogue of Aharonov-Bohm effect by Ford and Vilenkin in the linearized regime. Employing the $1+3$ formulation of spacetime decomposition we study the same effect in the context of full Einstein field equations for stationary spacetimes. Applying our approach to static tube-like and cylindrical distributions of dust not only we recover their result but also obtain an extra term which is interpreted to be representing the classical version of the Colella-Overhauser-Werner effect (the COW experiment). 
\end{abstract}
\maketitle


\section{Introduction}
Aharonov and Bohm in their celebrated paper have shown that the interference pattern of an electrically charged ensemble of particles traveling on  a closed path in a region where the electromagnetic fields are absent can be affected by the electric and magnetic fields in the region from which they are excluded \cite{AB}. This result is usually interpreted as a manifestation of non-locality of electromagnetism or as a direct physical effect attributed to the electromagnetic four-potential. Looking for gravitational analogues, physical effects analogous to the electromagnetic Aharonov-Bohm have been considered by many authors through different approaches and perspectives \cite{Dowk,Kraus,Asht,ford,stachel,Aud,Burg}.\\
Basically two different versions of gravitational Aharonov-Bohm effect are discussed in the literature which could be attributed to the spacetime under consideration being static or stationary. In the first version arising in stationary spacetimes possessing gravitomagnetic field, use is made of the fact that  the gravitational field couples to all other fields. In this case taking advantage of analogy between (weak) gravitational fields and electromagnetism, it is shown, by solving the Klein-Gordon equation, that energy spectrum of a scalar particle in the region where
the gravitomagnetic field vanishes depends on the gravitomagnetic flux in the region from which it is excluded \cite{bez1989, harris}. Other studies following essentially the same approach but in different contexts could be found in \cite{bez1991,jensen}.\\
Also loop space variables in the gravitational field of a slowly rotating long cylindrical shell as well as a static cosmic string are being employed to investigate gravitational analogues of the Aharonov-Bohm effect \cite{bez1987,bez2004}. It is shown that when one parallel transports a vector around a rotating source it acquires a gravitationally induced phase shift which is proportional to the angular momentum of the source as a local quantity and this happens despite the fact that the Riemann tensor dose not depend on the angular momentum.\\
Closely related to the above version of the effect, Stachel considers globally stationary but locally static spacetimes and solves the eikonal equation in such backgrounds \cite{stachel}. He shows that the eikonal function is proportional to the period of a one-form which in the 1+3 (or threading) formulation of spacetime decomposition is called the {\it threading vector}. It can be shown that the period is nothing but the gravitomagnetic flux defined in the same formulation of spacetime decomposition. The important point about Stachel's approach is that the dependence of the phase shift on the rotation of the source is a classical result and could be obtained without an appeal to the weak field approximation.\\
In the present paper we turn our attention to the second version of the effect investigated by Ford and Vilenkin \cite{ford}. Since the role of potentials and fields in the electromagnetism are played in general relativity by the metric and Riemann curvature tensors respectively, one can study physical effects originated from the regions of non-zero curvature on particles moving in a region where curvature vanishes. In other words in this case there is no need for the notion of a gravitomagnetic field to produce the effect and it could even arise in a static spacetime. For this purpose they consider  a tube-like distribution of matter producing a spacetime with two Killing vectors ${\bf \xi} = \xi^a \partial_a$ and ${\bf \eta} = \eta^a \partial_a$ \footnote{Apart from the equations in which the range of the indices are given explicitly, the Latin indices run from 0 to 3 while Greek indices run from 1 to 3.}( which are timelike and spacelike respectively) and asymptotically flat in the direction perpendicular to the tube. Then they make use of a form of Gauss-Bonnet theorem  stating that if a vector is parallel transported along a closed curve $C$ in a 2-surface $S$ it will not go back to itself but undergo a rotation by an angle $\alpha$  given by the following area integral 

\begin{equation}
\alpha=\int_{S_0} K da \label{K},
\end{equation}

where $K$ is the Gaussian curvature of the 2-surface and $S_0$ is the subsurface of $S$ enclosed by the curve. Now assuming that the curve $C$ lies in the asymptotically flat region then the particle moving on this curvature-less region still feels the effect from non-zero-curvature region, hence a version of
gravitational Aharonov-Bohm effect. To give an explicit example, they have applied their formulation to the case of a tube-like dust source using weak field approximation to calculate the  Gaussian curvature in terms of the components of the energy-momentum tensor of the source.\\
In this article we employ the same approach but in the context of the 1+3-decomposition of spacetimes which leads to the quasi-Maxwell form of the Einstein field equations and the so called {\it gravitoelectromagnetism}. This decomposition introduces a differentiable 3-manifold with a metric element
prescribing spatial distances in a given stationary spacetime. This in turn allows one to express the  Gaussian curvature of a 2-surface in terms of the components of the energy-momentum tensor of the source and  the {\it gravitoelectric} field of the underlying spacetime without using the weak field approximation. Outline of the paper is as follows: first we  introduce briefly the 1+3 decomposition of a stationary spacetime in section II and then, using the projection tensor defined on that basis, in section III we calculate an expression for the Gaussian curvature in terms of the 3-dimensional Ricci scalar and Ricci tensor. In section IV we apply this expression to different solutions including  static dust solutions and spacetime of a cosmic string. In the last section the results are summarized.

\section{1+3 (threading) formulation of spacetime decomposition}
Considering the light propagation in a  stationary spacetime between two nearby spatial points $x^\alpha$ and $x^\alpha + dx^\alpha $, the spacetime metric could be written in the following general form \cite{Landau}
\begin{equation}\label{metric1}
ds^2 = d\tau_{syn}^2 - dl^2 = h(dx^0 - A_\alpha dx^\alpha)^2-\gamma_{\alpha\beta} dx^\alpha dx^\beta,
\end{equation}
where $h \equiv g_{00}$, $A_\alpha = -\dfrac{g_{0\alpha}}{g_{00}}$ and  
\begin{equation*}
\gamma_{\alpha\beta}= -g_{\alpha\beta}+\dfrac{g_{0\alpha}g_{0\beta}}{g_{00}},
\end{equation*}
is the spatial metric. In this so-called $1+3$ formulation of spacetime decomposition $d\tau_{syn} = \sqrt{h}(dx^0-A_\alpha dx^\alpha)$ is the infinitesimal interval of {\it synchronized proper time} and $dl^2$ is the infinitesimal spatial distance between the two events. The spacetime being stationary ($g_{ij}\neq g_{ij}(x^0)$) will allow one to define distance along a curve in a finite region of the spacetime by the integral $ \int dl$.
In a more rigorous mathematical language, a stationary spacetime $({\cal M}, g_{ab})$ is decomposed into spatial and temporal sections  by a congruence of timelike curves generated by the timelike Killing vector field $\xi^a$ of the spacetime. This is achieved through the introduction of the following projection tensor 
\begin{equation*}
h_{ab}=-g_{ab}+u_a u_b,
\end{equation*}
where $u^a$ is the normalized tangent vector to the timelike curves
\begin{equation*}
u^a= \dfrac{\xi^a}{|\xi|}\;\;\; ; \;\;\; |\xi| = (\xi_a\xi^a)^{1/2}.
\end{equation*}
Using a coordinate system (denoted by the sign $\doteq$) in which $\xi^a\doteq (1, 0, 0, 0) $ ( and $\xi_a \doteq (g_{00}, g_{\alpha 0}) $ ), i.e adapted to the timelike Killing vector field, then
\begin{equation*}
g_{00}\doteq |\xi|^2=\xi_0 \;\;\; ; \;\;\; A_\alpha \doteq -\frac{\xi_\alpha}{\xi_0}\;\;\; ; \;\;\; h_{00} \doteq 0, 
\end{equation*}
and 
\begin{equation*}
\gamma_{\alpha\beta} \doteq h_{\alpha\beta} \doteq - g_{\alpha\beta} + \frac{1}{\xi_0}{\xi_\alpha}{\xi_\beta}, 
\end{equation*}
so that the spacetime metric will take the following form 
\begin{equation*}
ds^2 \doteq \xi_0(dx^0 + \frac{\xi_\alpha}{\xi_0} dx^\alpha)^2-h_{\alpha\beta} dx^\alpha dx^\beta.
\end{equation*}
It should be noted that $\Sigma_3$ is a differentiable 3-manifold but not a hypersurface in $\cal M$. 
Indeed it is called the quotient space $\frac{\cal M}{G_1}$ where $G_1$ is the one dimensional group of 
motions generated by the timelike Killing vector field of the spacetime $\cal M$ \cite{Steph}.\\
Working in the general coordinate system of (\ref{metric1}), in the 1+3-decomposition, {\it gravitoelectric} 
and {\it gravitomagnetic} fields are defined in terms of the derivatives of the metric components as follows \cite{LBNZ}
\begin{equation*}
\textbf{E}_g=-\dfrac{\nabla h}{2h} \doteq -\nabla \ln |\xi|,	~~~\textbf{B}_g=\nabla \times \textbf{A}.
\end{equation*}
In terms of the above fields, Einstein field equations for a perfect fluid source could be rewritten 
in the following {\it quasi-Maxwell} form \cite{LBNZ}
\begin{gather}
\nabla \times ~\textbf{E}_g=0, ~~~\nabla \cdot  \textbf{B}_g=0 \\
\nabla \cdot \textbf{E}_g= 1/2 h B^2_g+E^2_g - 8\pi[\dfrac{p+\rho}{1-v^2}-\dfrac{\rho-p}{2}] \label{r00}\\  \label{r01}
\nabla \times  (\sqrt{h}\textbf{B}_g)=2 \textbf{E}_g \times (\sqrt{h}\textbf{B}_g)-16\pi[\dfrac{p+\rho}{1-v^2} \textbf{v}] \\
{^{(3)}}R^{\mu\nu}=-{E}_g^{\mu;\nu}+h(B_g^\mu B_g^\nu - B_g^2 \gamma^{\mu\nu})+ {E}_g^\mu E_g^\nu+ 8\pi[\dfrac{p+\rho}{1-v^2}v^\mu v^\nu+\dfrac{\rho-p}{2}\gamma^{\mu\nu}]. \label{3ricci}
\end{gather}
The following points need to be mentioned with respect to the above equations:\\
1- ${^{(3)}}R^{\mu\nu} $ is the 3-dimensional Ricci tensor of the 3-space $\Sigma_3$ constructed from the 3-dimensional metric $\gamma_{\alpha\beta}$ 
in the same way that usual 4-dimensional Ricci tensor $R^{ab}$ is made out of $g_{ab}$.\\
2-In the above equations all the differential operations are defined in the 3-space $\Sigma_3$ with metric $\gamma_{\alpha\beta}$ \cite{Landau, LBNZ}.\\
3-The 3-velocities $v^\mu$ are defined with respect to the {\it synchronized proper time} i.e $v^\mu = \dfrac{dx^\mu}{d\tau_{syn.}}$.\\
Another feature in the above equations, which will be employed later, is the simple fact that by equation (\ref{r01}) a non-vacuum, static ($B_g \doteq 0$) solution produced by a perfect fluid, in general has to be in the comoving frame ($ v^\mu \doteq 0 $) with respect to the dust particles. In other words,  the same frame in which the spacetime has no cross terms is also comoving \footnote{ An obvious exception is the case where the equation of state is that of dark energy/cosmological constant, i.e $p = -\rho$.}.
\section{Gaussian Curvature in 1+3 formalism and gravitational Aharonov-Bohm Effect}
After introducing their formulation of gravitational Aharonov-Bohm effect, based on \eqref{K}, Ford and Vilenkin faced the problem of expressing Gaussian curvature in terms of the four-dimensional Ricci tensor so that, through Einstein field equations, to be able to calculate the phase shift \eqref{K} in terms of local quantities such as mass and angular momentum \cite{ford}. To do so they have appealed to the weak field limit to express $K$ in terms of the energy density of a dust source. In what follows, using the $1+3$ formulation and quasi-Maxwell form of the Einstein field equations and without employing the weak field limit, we show that the Gaussian curvature can be expressed in terms of the gravitoelectromagnetic fields of the source and consequently in terms of the components of its energy-momentum tensor.\\
To follow Ford and Vilenkin we consider a stationary space-time  corresponding to a tube-like distribution of matter with two Killing vectors ${\bf \xi} = \xi^a \partial_a$ and ${\bf \eta} = \eta^a \partial_a$ which are timelike and spacelike respectively. Let $S$ be a 2-surface in $\Sigma_3 $  orthogonal to the two Killing vectors and $C$ a closed curve in the 2-surface which may or may not encircle the tube containing the source. Now according to the relation \eqref{K} if a vector is parallel transported around $C$  it will acquire a rotation angle $\alpha$ due to the non-zero curvature region in the 2-surface $S$. To relate the Gaussian curvature of the 2-surface to the gravitoelectromagnetic fields of the underlying spacetime we introduce, using the spacelike Killing vector $\bf \eta$, a projection tensor from $\Sigma_3$ to $S$ as follows
 \begin{equation}
\tilde{h}_{\alpha\beta}=\gamma_{\alpha \beta}-n_\alpha n_\beta \;\; ; \;\; \alpha, \beta = 1,2,3 \label{proj2}
\end{equation}
where now $n_\alpha=\dfrac{\eta_\alpha}{|\eta|}$ is the unit vector normal to $S$.\\
Choosing a preferred coordinate system in which $\eta \doteq \partial_z$, i.e  $\eta^\alpha$ takes the following form
\begin{equation*}
\eta^\alpha\doteq(0,1,0), ~~~~ x^1=r, ~x^2=z, ~x^3=\phi ,
\end{equation*}
for two-surface $S$ with metric \footnote{This is so because the spatial metric $\gamma_{\alpha\beta}$ is independent of $t$ and $z$.}
\begin{equation}
\tilde{g}_{ij}=\gamma_{ij} \;\;\; ; \;\;\; i,j = 1,3
\end{equation}
the Gaussian curvature $K$ is given by 
\begin{equation}
K=\dfrac{1}{2}~ \tilde{g}^{ik} \tilde{g}^{jl}{^{(2)}}R_{ijkl} = \dfrac{1}{2}{^{(2)}}R \;\;\; ; \;\;\; i,j,k,l = 1,3
\end{equation}
Upper left indices indicate the dimension of the space for which the geometric entity is computed so in the above equation the upper left index $(2)$ shows that the Riemann tensor is calculated for the 2-surface $S$.\\
Since $\bf \xi$ is a Killing vector of the $\Sigma_3$ space and orthogonal to the 2-surface $S$ one can show that the extrinsic curvature of the 2-surface  is zero  and hence by Gauss-Codazzi equation we have ${^{(2)}}R_{kilj}={}^{(3)}R_{kilj}~~ (i,j,k,l= 1,3)$. This could be used in turn to calculate the two-dimensional Ricci tensor as follows
\begin{equation*}
{^{(2)}}R_{ij} = \tilde{g}^{kl} {^{(2)}}R_{kilj}= \tilde{h} ^{\alpha\beta} {^{(3)}}R_{\alpha i\beta j}~~;~~ i,j= 1,3 \\
\end{equation*}
where in the last equality by replacing for the projection tensor (\ref{proj2}) we end up with
\begin{equation}
{^{(2)}}R_{ij}={^{(3)}}R_{ij}-\dfrac{\eta^\alpha \eta^\beta}{|\eta|^2} {^{(3)}}R_{\alpha i \beta j}~~;~~ i,j= 1,3
\end{equation}
Therefore the two-dimensional Ricci scalar of the 2-surface $S$ with coordinates $\lbrace x^1, x^3\rbrace$ and metric $\tilde{g}^{ij}$ is given by
\begin{align*}
{^{(2)}}R &= \tilde{g}^{ij} {^{(2)}}R_{ij} \\
&=\gamma^{ij}( {^{(3)}}R_{ij}-\dfrac{\eta^{\alpha}\eta^\beta}{|\eta|^2} {^{(3)}}R_{\alpha i \beta j})~~;~~ i,j= 1,3
\end{align*}
where use is made of the fact that $\tilde{g}^{ij}=\gamma^{ij}$. Now applying the relation $ {^{(3)}}R_{\alpha\beta}=\gamma^{\mu\nu} {^{(3)}}R_{\alpha \mu \beta \nu}$ to the above equation we obtain the following relation for the Gaussian curvature of the 2-surface in terms of the 3-dimensional Ricci scalar and Ricci tensor of the 3-space $\Sigma_3$
\begin{equation}
{^{(2)}}R = 2 K = {^{(3)}}R-2 \dfrac{\eta^\alpha \eta^\beta}{|\eta|^2} {^{(3)}}R_{\alpha\beta}. \label{r2}
\end{equation}
Up to now we have considered stationary tube-like spacetimes which possess two killing vectors but it is obvious that the above formulation could also be applied to cylindrically symmetric spacetimes which have an extra Killing vector ${\bf \zeta} \doteq {{\partial}}_\phi$. In the next two subsections first we apply the above relation to a static dust solution and then to  cylindrically symmetric spacetimes to study a gravitational analogue of the Aharonov-Bohm effect.
\subsection{Static tube-like distribution of dust : Aharonov-Bohm effect vs the Colella-Overhauser-Werner effect}
As it is clear from its derivation, relation \eqref{r2} can be applied to any tube-like {\it stationary spacetime} with a given energy-momentum tensor, but to compare our results with those of Ford and Vilenkin \cite{ford} first  we apply it to a static tube-like distribution of dust given by the following general metric 
\begin{equation}\label{met}
ds^2=g_{tt} dt^2-g_{rr} dr^2-g_{zz} dz^2-g_{\phi\phi} d\phi^2,
\end{equation}
where all the metric components are functions of $r$ and $\phi$. For such a spacetime from contraction of equation \eqref{3ricci} (with $B_g=0$, $p=0$ and $v=0$) we arrive at
\begin{equation}
{^{(3)}} R = -\nabla \cdot \textbf{E}_g + E^2_g + 12 \pi\rho \label{d1} = 16 \pi\rho,
\end{equation}
where we have used \eqref{r00} in the last step. For the second term in the right hand side of \eqref{r2} we obtain
\begin{equation}
2 \dfrac{\eta^\alpha \eta^\beta}{|\eta|^2} {^{(3)}}R_{\alpha\beta}=8 \pi \rho-\textbf{E}_g \cdot \nabla ln|\eta|^2,\label{d2}
\end{equation}
after substituting equations \eqref{d1} and \eqref{d2} in \eqref{r2}, the Gaussian curvature for static dust yields
\begin{equation}
K = 4 \pi \rho + \textbf{E}_g \cdot \nabla ln|\eta|. \label{kd}
\end{equation}
The first term in \eqref{kd} is the exact same term that was obtained by Ford and Vilenkin \cite{ford} for a tube-like distribution of dust in the weak field regime, but the second term which depends on the gravitoelectric field of the source and generally non-zero is absent in their approach. Now taking the spacetime \eqref{met} to be asymptotically flat ( where $E_g \rightarrow 0$ as $r \rightarrow \infty$), by equation \eqref{K}, a particle parallel transported along a closed path $C$ in this zero-curvature region will nevertheless be affected by the non-zero curvature in the interior region.\\
Since the second term is proportional to the gravitoelectric field, when integrated over a closed path, it reminds one of the Colella-Overhauser-Werner (COW) experiment/effect \cite{Col}. This term arising from a purely geometrical consideration, shows that the gravitationally induced phase shift has a classical origin \footnote{This is previously claimed by Mannheim \cite{mann} on other grounds.}. The presence of this term shows that even when the closed path of a particle does not encircle the source, there still will be an effect in the interference pattern resulting from the presence of the gravitoelectric field. To investigate in more detail the effect of this extra term, in the next subsection we consider cylindrically symmetric dust spacetimes/sources  and their matching to exterior (vacuum) solutions.
\subsection{Static cylindrically symmetric counter rotating dust}
Consider matching of a cylindrically symmetric static dust solution, at a given radius $R$, to an exterior (vacuum) asymptotically flat static solution, both of the general type \eqref{met} but now all metric components are functions only of $r$. As an explicit example one can think of the interior solution introduced by Teixeira and Som \cite{Teix}, representing counter rotating dust particles with net zero angular momentum and the following energy-momentum tensor\footnote{In what follows we are not concerned with the explicit form of their solution and the interested reader could refer to \cite{Teix}.} 
\begin{equation}\label{pf}
T^a_b = \frac{1}{2}\rho (u_a u^b + v_a v^b),
\end{equation}
where $u^a \doteq (u^0, 0, 0, \omega)$ and $v^a \doteq (u^0, 0, 0, -\omega)$ are the four velocities of the counter rotating particles with $u^au_a = v^a v_a = 1$. This interior spacetime is matched to the well-known exterior Levi-Civita metric \cite{Gri}
\begin{equation}
ds^2=r^{4\sigma}dt^2 - r^{-4\sigma}\left[r^{8\sigma^2}(d r^2+ B^2dz^2)+ C^2r^2d\phi^2 \right],\label{lc}
\end{equation}
in which $B$ and $C$ are scaling parameters and $\sigma$, for small values, could be interpreted as the effective gravitational mass per unit proper length \footnote{ Obviously in the process of matching the two solutions it is is expected that $\sigma$ to be related to the parameters of the static dust solution. For interpretation of the parameters of the Levi-Civita solution and possible matchings of interior and exterior solutions refer to \cite{Gri} and references therein.}. Now two cases could be considered:\\
I-The closed path of the parallel transported particle $C$ encircles the non-vacuum cylindrical region (Fig.1) so that the rotation angle \eqref{K}
is given by
\begin{equation}
\alpha=\int_{S_0=S^{ext}_0 \cup S^{int}_0} K da = \int_{S^{ext}_0} (K~da)^{ext} + \int_{S^{int}_0} (K~da)^{int},\label{case1}
\end{equation}
where the upper indices show the region of spacetime in which the quantities are calculated.\\
Now repeating the calculation leading to \eqref{kd}, now with the energy-momentum tensor of the perfect fluid given by \eqref{pf} we end up with,
\begin{equation}\label{v2}
K^{int} = \frac{4 \pi \rho}{1 - v^2}  + (\textbf{E}_g \cdot \nabla ln|\eta|)^{int}, 
\end{equation}
for the Gaussian curvature of the interior region where $v^2$ is the squared norm of the three velocity of the dust particles defined by $v^\mu = \frac {dx^\mu}{d\tau}$ ($\tau$ being the proper time in the corresponding static spacetime). Indeed using the relation between the components of this three-velocity and the corresponding four-velocity $u^\mu = \frac{dx^a}{ds}$ \cite{Landau} one can obtain the following relation
\begin{equation}
v^2 = \frac{g_{\phi\phi} \omega^2}{g_{\phi\phi} \omega^2 -1}.
\end{equation}
\begin{figure}
\begin{center}
\includegraphics[angle=0,scale=0.6]{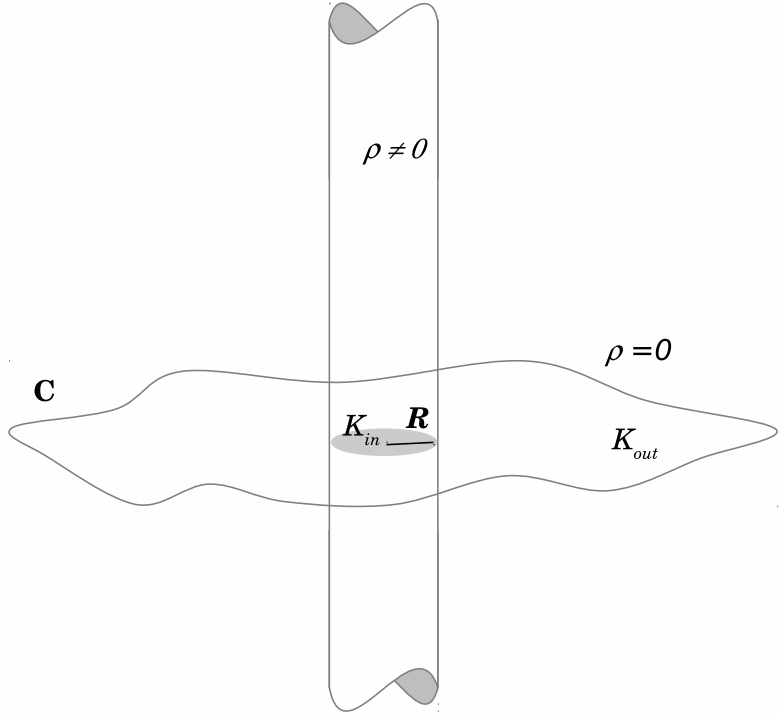}
\caption{Closed path $C$ of a particle encircling the non-vacuum cylindrical region at the asymptotically flat region.}
\end{center}
\end{figure}
\begin{figure}
\begin{center}
\includegraphics[angle=0,scale=0.6]{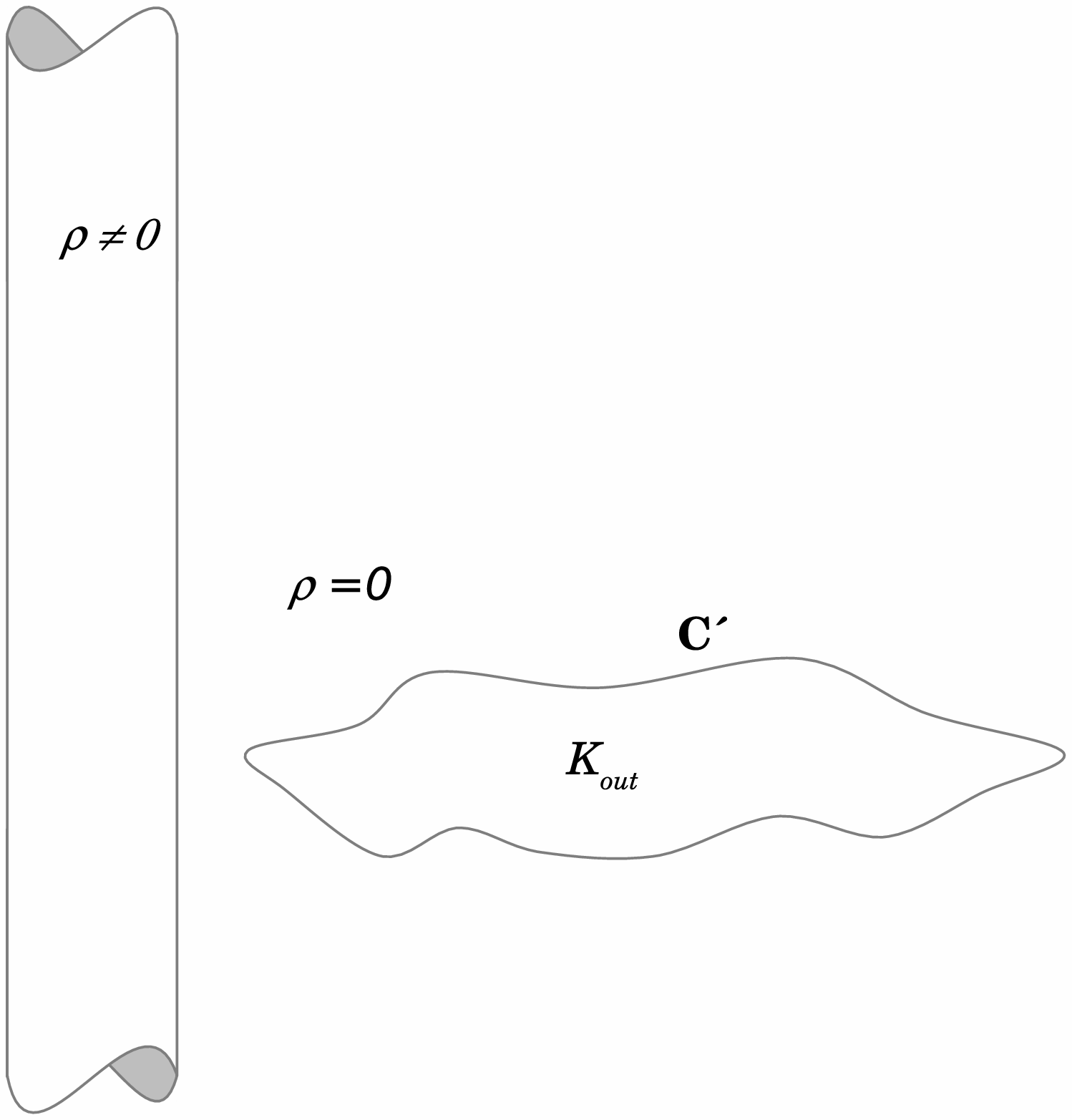}
\caption{Closed path $C'$ of a particle  which does not encircle the non-vacuum cylindrical region. }
\end{center}
\end{figure}
Equation \eqref{v2} is an interesting result showing that not only the density of the particles but also norm of their velocity, both as local quantities, have global effects. The presence of $v^2$ is a manifestation of the fact that the kinetic energy of the particles gravitates even though their net angular momentum is zero.\\
Obviously for the exterior (vacuum) static solution ( $\rho = 0$) we end up with the following relation for the Gaussian curvature
\begin{equation}
K^{ext} =  (\textbf{E}_g \cdot \nabla ln|\eta|)^{ext}.
\end{equation}
The overall result in the arrangement of Fig.1, as in the case of a simple static dust solution, for a path $C$ in the asymptotically (locally) flat region, is a rotation angle \eqref{case1} which is a manifestation of a gravitational Aharonov-Bohm effect.\\
II-In this case the closed path of the parallel transported particle $C'$  does not encircle the non-vacuum cylindrical region (Fig.2) so that the rotation angle \eqref{K} is given by
\begin{equation}
\alpha=\int_{S^{ext}_0} (K~da)^{ext} = \int ({\bf E}_g \cdot \nabla ln|\eta|~ da)^{ext},\label{case2}
\end{equation}
which is obviously of the COW-type effect leading to gravitationally induced phase shift on the transported particle. 
In other words the Gaussian curvature  having a non-zero contribution from the second term in (\ref{kd}) represents the COW-type effect for a cylindrical gravitational field represented by the above exterior metric. Indeed in this approach, both gravitational Aharonov-Bohm effect and the COW effect/experiment, for cylindrically symmetric spacetimes, were treated as different phase shifts under the same formulation.

\subsection{Deficit angle of a cosmic string}
As another somewhat trivial example of the application of the formalism introduced in section III, we turn our attention to the space-time of a cosmic string that seems to be naturally adapted to the investigations on gravitational Aharonov-Bohm effect. Space-time metric of a cosmic string
is given by setting $\sigma = \frac{1}{2}$ in \eqref{lc} (with the rescaling parameter $B$ removed) \cite{vilenkin}
\begin{equation}
ds^2=dt^2-d r^2-dz^2-C^2 r^2d\phi^2 \label{scs}.
\end{equation}
The corresponding energy-momentum tensor is given by \cite{Linet}
\begin{equation}
T^t_t=T^z_z=\mu \dfrac{\delta({\bf r})}{\sqrt{\tilde {g}}},  ~~~~ T^\rho_\rho=T^\phi_\phi=0, \label{emscs}
\end{equation}
where $\mu$ is the linear mass density and $\tilde {g}$ is the determinant of the metric on the $t=\rm const.$ and $z=\rm const.$ 2-surface $S_0$. To compute the Gaussian curvature we rewrite the last of quasi-Maxwell equations, equation \eqref{3ricci}, for the above  
spacetime metric and energy-momentum tensor as follows (with $E_g = B_g =0$ )
\begin{equation}
{^{(3)}}R^{\mu\nu}=8\pi(T^{\mu\nu} - \frac{1}{2} g^{\mu\nu} T). \label{nr3}
\end{equation}
Contraction with $\gamma_{\mu\nu}$ gives the 3-dimensional Ricci scalar as follows
\begin{equation}
{^{(3)}}R=16\pi(\mu \dfrac{\delta({\bf r})}{\sqrt{\tilde {g}}}). 
\end{equation}
On the other hand it is an easy task to see that in the adapted coordinate system (in which $\eta^\mu \doteq (0,1,0)$ ) the second term in the right hand side of relation \eqref{r2} vanishes so that the Gaussian curvature in this case is given by
 \begin{equation}
K=\frac{1}{2} {^{(3)}}R=8 \pi(\mu \dfrac{\delta({\bf r})}{\sqrt{\tilde g}}). 
\end{equation}
After performing the area integral of $K$ over two-surface $S_0$ we have the rotation angle $\alpha=8 \pi \mu$. Now employing the  the Gauss-Bonnet theorem 
\begin{equation}
\int_S K ~da = 2\pi\chi(S)-\int_{\partial S} k_g dl, \label{gb}
\end{equation}
to the above case we can find the deficit angle $2\pi\delta$ corresponding to a cosmic string. To do so we take $\chi(S)=1$ and $ k_g=\frac{1}{r}$ for Euler characteristic and geodesic curvature of the corresponding 2-surface $S_0$ and its encircling path ${\partial S_0}$ respectively i.e,
\begin{equation}
\int_{S_0} 8\pi \mu {\delta({\bf r})} ~d^2{\bf r} = 2\pi-\int_{\partial S_0} \frac{1}{r} C r d\phi, \label{gb1}
\end{equation}
leading to the scaling factor $C \equiv 1 - \delta = 1-4\mu$ of a cosmic string in terms of its linear mass density.

\section{Conclusion}
With the help of a projection tensor defined in the context of $1+3$ formulation of spacetime decomposition we introduced an expression for the Gaussian curvature of a 2-surface in terms of the  Ricci scalar and Ricci tensor of the corresponding 3-space $\Sigma_3$. On the other hand using the quasi-Maxwell form of the Einstein field equations we have related the components of the energy-momentum tensor of a source and the grvitoelectromagnetic fields of the underlying spacetime to the 3-dimensional Ricci tensor of $\Sigma_3$. This enabled us to relate the Gaussian curvature of a 2-surfaces in $\Sigma_3$ to local quantities such as the energy density of the source and velocity norm of its constituent particles. The expression is then applied to the case of a tube-like perfect fluid source to investigate a version of gravitational Aharonov-Bohm effect introduced by Ford and Vilenkin. While they have applied their formalism to a static dust solution in the weak field regime here we have shown that our formulation allows one to apply it to {\it stationary spacetimes} and without  appeal to the weak field regime. In doing so not only we have recovered, for a static dust solution,  the same result as that of Ford and Vilenkin but also obtained, as an intersting by-product, the appearance of a phase shift related to the presence of the gravitomagnetic field of the underlying spacetime. It is shown that this phase shift could be interpreted as the {\it classical} version of the well known COW (experiment) effect. This was shown explicitly in the context of a static cylindrically symmetric vacuum solution matched to an interior static dust solution. Indeed applying our formulation to the case of a dust solution produced by counter-rotating particles we have arrived at a Gaussian curvature which is related not only to the energy density of the source but also to the norm of the particle velocities, despite the fact that their net angular momentum is zero. As another trivial example of the application of our formalism, also showing its consistency, deficit angle of a cosmic string was obtained using the Gauss-Bonnet theorem.



\end{document}